\documentstyle[12pt]{article}

\csname @addtoreset\endcsname{equation}{section}

\def\bseq{\begin{subequation}}  
\def\eseq{\end{subequation}}
\def\bsea{\begin{subeqnarray}}  
\def\esea{\end{subeqnarray}}

\def\bbar{b \kern-.40em /}

\newcommand{\bbox}{\lower.2ex\hbox{$\Box$}}

\newcommand{\beq}{\begin{equation}}
\newcommand{\eeq}{\end{equation}}
\newcommand{\bea}{\begin{eqnarray}}
\newcommand{\eea}{\end{eqnarray}}
\newcommand{\ena}{\end{eqnarray}}

\newcommand{\wt}{\tilde}
\renewcommand{\a}{\alpha}
\renewcommand{\b}{\beta}
\renewcommand{\c}{\gamma}
\renewcommand{\d}{\delta}
\newcommand{\th}{\theta}

\newcommand{\pa}{\partial}
\newcommand{\g}{\gamma}
\newcommand{\G}{\Gamma}

\newcommand{\D}{\Delta}
\newcommand{\e}{\epsilon}

\newcommand{\p}{\pi}

\renewcommand{\t}{\tau}

\newcommand{\r}{\right}
\newcommand{\nn}{\nonumber}
\def\II{\relax{\rm I\kern-.60em 1}}
\newcommand{\su}{s_1}
\newcommand{\sd}{s_2}
\newcommand{\st}{s_3}
\newcommand{\vb}{v_\beta}
\newcommand{\vg}{v_\gamma}
\newcommand{\Sh}[2]{\ {\rm Sinh} (#1 #2)}
\newcommand{\Ch}[2]{\ {\rm Cosh} (#1 #2)}
\newcommand{\sha}{\Sh{\su}{\va}}
\newcommand{\shb}{\Sh{\sd}{\vb}}
\newcommand{\shc}{\Sh{\st}{\vg}}
\newcommand{\cha}{\Ch{\su}{\va}}
\newcommand{\chb}{\Ch{\sd}{\vb}}
\newcommand{\chc}{\Ch{\st}{\vg}}
\newcommand{\shu}{\Sh{\su}{\va}}
\newcommand{\shd}{\Sh{\sd}{\va}}
\newcommand{\chu}{\Ch{\su}{\va}}
\newcommand{\chd}{\Ch{\sd}{\va}}

\newcommand{\va}{v_\a}

\begin{document}

\begin{titlepage}
\begin{flushright} IFUM--643--FT\\
\end{flushright}
\vfill
\begin{center}
{\LARGE\bf Three-graviton scattering and recoil effects\\
\vskip 3.mm
in M-atrix theory}\\
\vskip 27.mm  
{\large\bf  Andrea Refolli, Niccolo' Terzi and  Daniela Zanon  } \\
\vskip 5.mm
{\small
Dipartimento di Fisica dell'Universit\`a di Milano and\\
INFN, Sezione di Milano, via Celoria 16,
I-20133 Milano, Italy}\\
\end{center}
\vfill

\begin{center}
{\bf ABSTRACT}
\end{center}
\begin{quote}
We study the scattering of three gravitons in $M$-atrix theory at finite 
$N$. With a specific choice of the background we obtain the complete
result up to two loops. The contributions from three-body
forces agree with the ones presented in recent papers. We extend 
the calculation and evaluate the two--body exchanges as well. 
Such terms, somewhat difficult to isolate and compute,
had been neglected so far in the existing literature. 
 We show that the result we have obtained from $M$-atrix theory
precisely matches the result from one-particle reducible tree diagrams 
in eleven-dimensional supergravity .

\vfill     
\vskip 5.mm
 \hrule width 5.cm
\vskip 2.mm
{\small
\noindent e-mail: andrea.refolli@mi.infn.it\\
\noindent e-mail: terzi@pcteor1.mi.infn.it\\
\noindent e-mail: daniela.zanon@mi.infn.it}
\end{quote}
\begin{flushleft}
May 1999
\end{flushleft}
\end{titlepage}

\section{Introduction}
Since the introduction of the matrix model of $M$-theory \cite{BFSS}
much work and much progress have been made on the subject.
According to the original conjecture,
the degrees of freedom of $M$-theory in the infinite momentum frame are
contained in the dynamics of $N$ $D0$-branes in the $N\rightarrow \infty$
limit. Subsequently it was argued that $M$-theory with one of the 
lightlike coordinate compactified, the so called discrete light cone sector,
is in fact equivalent to the super Yang-Mills matrix model for finite $N$
 \cite{LS}, \cite{SS}.
Compelling tests of this proposal
have been the comparison of two-body \cite{BB,BBP} and
three-body \cite{giappo,FFI,ALL,DR} scattering.

We focus on the $M$-atrix theory at finite $N$ and explore further its 
correspondence with eleven dimensional supergravity. In particular we 
consider $N=3$ and compute graphs up to two-loops in $(0+1)$-dimensional
Yang-Mills. We find that they perfectly reproduce the three graviton 
scattering in supergravity, both in the direct three-body channel and in 
the two-body recoil exchange. The former, which corresponds to
one-particle irreducible diagrams in supergravity
, has been computed in the first paper of ref. \cite{giappo} 
and we confirm that result. 
We complete the two--loop effective action calculation and 
evaluate two--body exchanges as well.
We have found a systematic way to separate the light and
heavy matrix model degrees of freedom which allows to obtain the full
answer. We show that these contributions exactly match what
expected from the two-loop scattering of two $D0$-branes in $M$-atrix
theory \cite{BBP}.

We note that the recoil effects we have computed are in the context 
of an effective action calculation, i.e. two-loop 1P--irreducible diagrams in 
Matrix theory. In the second paper of  ref. \cite{giappo} a detailed
analysis of recoil interpreted as geodesic acceleration due to
one--loop corrections was given. These contributions correspond in Matrix 
theory to two--loop 1P--reducible diagrams.

In the next section we derive the explicit form of the gauged fixed action.
The various fields are decomposed in terms of components on a $U(3)$ basis 
of hermitean matrices. The classical background is fixed with the three
$D$-particles having relative velocities parallel to each others, orthogonal 
to the corresponding
relative displacements. The Feynman rules are easily derived. In section 3
we give the two-loop effective action, leaving most of the 
technical details in the Appendices. The result is analyzed keeping separate
the two types of contributions mentioned above, i.e. terms which depend on 
two distinct relative velocities (three-body interaction)
 and terms in which only one relative 
velocity appears (recoil).
The last section contains a summary of our results. 
We have performed part of the calculations  with the help of Mathematica.

\section{ The action}
As anticipated in the introduction, the matrix model is simply obtained by
reducing $(9+1)$-dimensional $U(N)$ super Yang-Mills \cite{YM10}
to $(0+1)$ dimensions. 
This theory describes a system of $N$ $D0$-branes \cite{PW}
in terms of nine 
 bosonic fields $X_i$ and of sixteen fermionic superpartners
$\theta$, which are spinors under $SO(9)$. The euclidean action is given by
\bea
S&=& {\rm Tr} \int d\tau \left \{ \left(D_{\tau}{X}_i\right)^2 
- \frac{g}{2}[X_k,X_j][X_k,X_j] \right. \nn\\
&&~~~~~~~~~~~~~~~~~~~\left.+ \th^T D_{\tau}{\th}
- \sqrt{g} \th^T \g^k [X_k,\th] \right\}
\label{action0}
\eea
where we have denoted by $g$ the Yang-Mills coupling constant
and by $\g^i$ nine
real, symmetric gamma matrices satisfying $\{\g^i,\g^j\}=2\d^{ij}$.
The covariant derivative is defined by
\beq
D_\tau=\pa_\tau -i \sqrt{g}[A,~~~]
\label{covder}
\eeq
The fields $X_i$, $\theta$ and $A$ are $N \times N$ hermitean
matrices of $U(N)$, with $i,j,k=1,2,\dots,9$.

Being interested in quantum, perturbative calculations it is 
convenient to use the background field method, which allows to maintain
explicit the gauge invariance of the result. To this end one expands the
action (\ref{action0}) around
a classical background field configuration $B_i$, setting
$X_i\rightarrow X_i+B_i$. After gauge-fixing and trivial rescaling of the fields, the complete action is 
\bea
S&=& {\rm Tr} \int d\tau \left \{ \left(\pa_{\tau}{X}_i\right)^2 
- [B_k,X_j]^2 - 2 \sqrt{g}
[B_k,X_j][X_k,X_j]- \frac{g}{2}[X_k,X_j][X_k,X_j] 
\right. \nonumber \\
&&~~+ \pa_{\tau}{A}^2 - [A,B_k]^2 - 4 i\pa_{\tau}{B}_k [A,X_k]- 
2 i\sqrt{g} \pa_{\tau}{X}_k [A,X_k]+ 2 \sqrt{g} [A,B_k][X_k,A]  \nonumber \\
 &&~~- g[A,X_k]^2 + \th^T \pa_{\tau}{\th} - i \sqrt{g}  \th^T [A,\th] 
- \sqrt{g} \th^T \g^k [X_k,\th] -   \th^T \g^k [B_k,\th] 
 \nonumber \\
&&~~ \left. - 2 \tilde G \pa_\tau^2 G  -2i\sqrt{g}
 \pa_{\tau}{\tilde G} [A,G]  + 2 \tilde G \left[ B_k, [B_k,G] -  
\sqrt{g} [G,X_k] \right]
\right \}
\label{action1}
\eea
Here $X_i$, $A$ and $\theta$ are the quantum fluctuations, $G$ and 
$\tilde{G}$ 
are the ghosts, while $B_k$ is the external background. 

Since we want to extract results to be compared to the scattering
of three gravitons in supergravity, the minimal choice for the 
Yang-Mills gauge group
that allows to describe the interaction of three $D0$-branes, is $U(3)$.
In Appendix  A we explicitly 
give the Cartan 
basis $H^1$, $H^2$, $E_\a$ (where $\a=\pm \a^1,\pm\a^2,\pm\a^3$ are the 
roots), and the commutator algebra.

 Using such a basis every matrix field is decomposed into components as
\bea
X_k&\equiv& X_k^a H_a + X_k^\a E_\a + X_k^{*\a } E_{- \a}\nonumber\\ 
A& \equiv& A^a H_a + A^\a E_\a + A^{*\a } E_{- \a} \nonumber\\
\theta&\equiv& \theta^a H_a + \theta^\a E_\a + \theta^{*\a } E_{- \a}
\nonumber\\ 
G&\equiv& G^a H_a + G^\a E_\a + G^{*\a } E_{- \a} \nonumber\\
\tilde{G}&\equiv& \tilde{G}^a H_a + \tilde{G}^{*\a} E_\a + 
\tilde{G}^\a  E_{- \a} 
\label{scompo}
\eea
In order to keep the notation simple we use a greek index $\a=1,2,3$ as a
suffix  to enumerate some of the component fields and the same index 
to indicate 
the corresponding
root on the matrices, i.e. $E_1 \equiv E_{\a^1}$, $E_2 \equiv E_{\a^2}$,
$E_3 \equiv E_{\a^3}$.
Similarly, in the following of
the paper the roots 
will be denoted with the same understanding: to be more explicit, if we write
$X^\a X^\b \a\cdot\b$  we mean for example when $\a=1$ and $\b=2$
$\to$ $X^1 X^2 \a^1\cdot\a^2$.
Notice in addition that the hermiticity condition gives e.g. 
$X_k^{-\a}=X_k^{*\a}$. 

Now we make a specific choice of the background configuration, i.e. 
straight line trajectories for the three particles. This amounts to have
$B_k$ in diagonal form with
\beq
B_k^\a=\tilde{v}^\a_k \tau +\tilde{b}^\a_k \qquad \qquad\qquad
\a=1,2,3
\label{backdiag}
\eeq
The free motion of the center of mass can be factored out and ignored
imposing
\beq
\sum_{\a=1}^{3} \tilde{v}^\a_k =0\qquad \qquad\qquad 
\sum_{\a=1}^{3}\tilde{b}^\a_k=0
\label{centermass}
\eeq
We introduce further simplifications restricting ourselves to the case of
parallel velocities for all three particles, e.g. along the $x_1$ axis,
and relative displacements transverse \cite{FFI}
\bea
&&\tilde{v}^\a_1 \neq 0\qquad,\qquad\tilde{v}^\a_k =0\qquad{\rm for}\quad k>1
\nn\\
&&\sum_{k=1}^{9}\tilde{v}^\a_k \tilde{b}^\a_k=0\qquad{\rm for}\quad \a=1,2,3
\label{simpleback}
\eea
Setting $\tilde{v}^\a_1=\tilde{v}_\a$ the background matrices become 
\beq
B_1=\left( \matrix{\tilde{v}_1 \t&0&0\cr 0&\tilde{v}_2 \t&0\cr 0&0&
\tilde{v}_3 \t}\right)
\qquad,\qquad 
B_k=\left( \matrix{\tilde{b}^1_k &0&0\cr 0&\tilde{b}^2_k &0\cr 0&0&
\tilde{b}^3_k}\right)
\qquad {\rm for} \quad k>1
\label{background}
\eeq
We define relative velocities
\beq
v_1= \tilde{v}_2-\tilde{v}_3 \qquad\qquad {\rm and} \qquad {\rm cyclic}
\label{relvel}
\eeq
and relative impact parameters
\beq
b^1_k= \tilde{b}^2_k-\tilde{b}^3_k \qquad\qquad {\rm and} \qquad {\rm cyclic}
\label{relimpact}
\eeq
In terms of these quantities, setting
\beq
R_k^\a=\sum_{a=1,2} \a_a {\rm Tr}\left( H^a B_k \right)
\eeq
we obtain
\beq
R_k^\a = \left \{ \matrix{
v_\a \tau ~~&{\rm if}~~ & k=1 \cr
b^\a_k ~~ &{\rm if}~~ & k > 1}\right. 
\label{moreback}
\eeq

Now we can go back to the action in (\ref{action1}) and perform the trace 
operation explicitly. Using the notation
\bea
(b^\a)^2 &\equiv&\sum_k b^\a_k b^\a_k \nonumber\\
(R^\a)^2 &\equiv&\sum_k R^\a_k R^\a_k\equiv v_\a^2\tau^2+(b^\a)^2
\label{moredef}
\eea
we obtain the following expressions: 

terms involving the $X$ fields:
\bea
S_X&=&  \int d\tau \left \{ X_k^a (- \pa^2_\tau) X_k^a +
 2 X_k^{*\a } (- \pa^2_\tau + \left( R^{\a}\right)^2)  X_k^\a 
- 2   \sqrt{g} \left[ \e^{\a \b \c}  R_k^{\c}( X_k^\a X_j^\b X_j^\c 
\right. \right. \nonumber \\
&&~~~~~~~~ \left. + X_k^{*\a} X_j^{*\b} X_j^{*\c } ) 
- 2  R_k^{\a} X_j^\a X_j^{*\a } X_k^a \a_a 
+ R_k^{\a} X_j^\a X_k^{*\a} X_j^a  \a_a+
    R_k^{\a} X_k^\a X_j^{*\a } X_j^a  \a_a \right ]  \nonumber \\
&&~~ - g \left [ -2 X_j^\a X_j^{*\a}(  X_k^a \a_a)^2
- 2  X_k^a \b_a X_j^\b X_k^\a X_j^\c  \e^{\a \b \c} +
2 X_k^a \b_a X_j^b \b_b X_j^\b X_k^{*\b }
\right.  \nonumber \\ 
&&~~~~~- 2  X_k^a \b_a X_j^{* \b} X_k^{* \a} X_j^{* \c}  \e^{\a \b \c} 
-  X_k^\a X_k^{*\eta } X_j^\b X_j^{*\rho}
\e^{\a \b \c}\e^{\eta \rho \c}  \nonumber \\ 
&&~~~~~\left. \left. 
+   X_k^\a X_j^{*\a }  X_k^\b X_j^{*\b } (\a \cdot \b) -
  X_k^\a X_j^{*\a}  X_j^\b X_k^{*\b}(\a \cdot \b) \right] \right \}
\label{actionbos}
\eea  

terms containing the gauge field $A$:
\bea
S_A&=&  \int d\tau \left \{ A^a (- \pa^2_\tau) A^a +
2 A^{*\a} (- \pa^2_\tau + (R^{\a})^2)  A^\a
+ 4 i \pa_{\tau}R_1^\a ( A^{*\a }X_1^\a - X_1^{*\a}  A^\a)
\right. \nonumber \\
&&~~ - 2 i  \sqrt{g} \left [ A^a \a_a X_k^\a \pa_{\tau}X_k^{*\a}
-  A^a \a_a \pa_{\tau}X_k^\a X_k^{*\a } +  A^\a \a_a 
\pa_{\tau}X_k^a X_k^{*\a }
-  A^\a \a_a X_k^a \pa_{\tau}X_k^{*\a} \right.\nonumber \\
&&~~~~~~ +\left.  A^{*\a } \a_a \pa_{\tau}X_k^\a X_k^a - 
 A^{*\a } \a_a \pa_{\tau}X_k^a X_k^\a
+ \e^{\a \b \c} A^\a X_k^\b \pa_{\tau}X_k^\c -
 \e^{\a \b \c} A^{*\a} X_k^{*\b } \pa_{\tau}X_k^{*\c } \right]
 \nonumber \\
&&~~-2\sqrt{g}\left[ R_k^{\a}A^b\a_b(X_k^{*\a}A^{\a} 
+X_k^{\a}A^{*\a})-2 R_k^{\a}A^{\a}A^{*\a}X_k^b\a_b \right.\nonumber \\
&&~~~~~~~~~+\left. \e^{\a\b\g} R_k^{\g}(X_k^{\a}A^{\b}A^{\g}
+X_k^{*\a}A^{*\b}A^{*\g})\right] \nonumber \\
&&- g \left[-2A^{\b}A^{*\b}(X_k^a\b_a)^2-
2\e^{\a\b\g}(X_k^a\b_a)(X_k^{\a}A^{\b}A^{\g}+
X_k^{*\a}A^{*\b}A^{*\g}) \right. \nonumber \\
&&~~~~~~  +2(X_k^a\b_a)(A^b\b_b)(A^{\b}X_k^{*\b}+A^{*\b}X_k^{\b}) -
2X_k^{*\b} X_k^{\b}(A^a\b_a)^2 \nonumber \\
&&~~~~~~ +2\e^{\a\b\g}(A^b\a_b)(X_k^{\a}X_k^{\g}A^{\b}+
X_k^{*\a}X_k^{*\g}A^{*\b}) -2\e^{\a\b\g}\e^{\eta\rho\g}
X_k^{\a}X_k^{*\eta}A^{\b}A^{*\rho} \nonumber \\
&&~~~~~~ \left.\left. +(\a \cdot \b) (X_k^{\a}X_k^{\b}A^{*\a}A^{*\b}+
X_k^{*\a}X_k^{*\b}A^{\a}A^{\b}-2 X_k^{\a}X_k^{*\b}A^{*\a}A^{\b}) \right]
\right\}
\label{actiongauge}
\eea

terms containing the fermions:
\bea
 S_{\theta} & = & \int{\rm d}\t\:\left\{\theta^{aT}\left(
\pa_{\t} \right)\theta^{a}+2\theta^{*\a T}\left(\pa_{\t}-
\g^i R_i^{\a}\right)\theta^{\a} \right. \nonumber \\
&& -2i\sqrt{g}\left[\theta^{a T}\a_a(\theta^{*\a}A^{\a}-
\theta^{\a}A^{*\a})  +\frac{1}{2} \e^{\a\b\g} (\theta^{*\a T}
\theta^{*\b}A^{*\g}-\theta^{\a T}\theta^{\b}A^{\g}) \right. \nonumber \\
&& \left. + \theta^{*\a T}\theta^{\a}A^a\a_a \right] - 2\sqrt{g}
\left[ \theta^{a T}\a_a \g^i (X_i^{\a}\theta^{*\a}-X_i^{*\a}\theta^{\a}) 
+ \theta^{*\a T}\g^i\theta^{\a}X_i^a\a_a \right. \nonumber \\ 
&&\left. \left. + \frac{1}{2} \e^{\a\b\g} (\theta^{*\a T} 
\g^i \theta^{*\b} X_i^{*\g}-\theta^{\a T} \g^i \theta^{\b} 
X_i^{\g}\right]\right\}
\label{actionfermion}
\eea

terms involving the ghosts:
\bea
S_{G}&=&\int{\rm d}\t \: \left\{2\wt{G}^{\a}\left(
-\pa_\tau^2+(R^{\a})^2\right)G^{\a}+2\wt{G}^{*\a}\left(-\pa_\tau^2+
(R^{\a})^2\right)G^{*\a}  \right. \nonumber \\
&& +2\wt{G}^a\left(-\pa_\tau^2\right) G^a +2\sqrt{g}\left[
-  R_k^{\a}\a_b(\wt{G}^{\a}X_k^{\a}+ \wt{G}^{*\a}X_k^{*\a})G^b \right. \nn\\
&&\left.  + R_k^{\a}\a_bX_k^b( \wt{G}^{\a} G^{\a}+
\wt{G}^{*\a} G^{*\a}) +\e^{\a\b\g}  R_k^{\g}(\wt{G}^{\g}G^{*\a}
X_k^{*\b}+\wt{G}^{*\g}G^{\a}X_k^{\b}) \right] \nonumber\\
&& +2i\sqrt{g}\left[\pa_{\tau}\wt{G}^a\a_a(G^{\a}A^{*\a}
-G^{*\a}A^{\a})+(\pa_{\tau}\wt{G}^{\a}A^{\a}-\pa_{\tau}\wt{G}^{*\a}
A^{*\a})G^a\a_a \right. \nonumber \\
&&\left. \left. -(\pa_{\tau}\wt{G}^{\a}G^{\a}-\pa_{\tau}\wt{G}^{*\a}
G^{*\a})A^a\a_a+\e^{\a\b\g}(\pa_{\tau}\wt{G}^{*\g}G^{\a}A^{\b}
-\pa_{\tau}\wt{G}^{\g}G^{*\a}A^{*\b})\right] \right\} \nn\\
&&~~~~~~~~~~
\label{actionghost} 
\eea
From the quadratic part of the action one can easily read the mass matrix
( diagonalizing further (\ref{actionbos}) and (\ref{actiongauge}))
and obtain the following particle content in the spectrum:
\begin{itemize}
\item 8 complex bosons with mass $(R^{\a})^2$ where
$\a=1,2,3$;
\item 1 complex boson with mass $(R^{\a})^2\pm 2v_{\a}$ where
$\a=1,2,3$;
\item 20 real, massless bosons;
\item 2 complex ghost fields with mass $(R^{\a})^2$ where
$\a=1,2,3$;
\item 2 real, massless ghosts;
\item 1  complex spinor with mass $\g^i  R_i^{\a}$ where
$\a=1,2,3$;
\item 2  Majorana, massless spinors.
\end{itemize}
Correspondingly we obtain the propagators of the various fields \cite{BB}. The bosonic
ones are given by
\bea
<X_i^a(\t)X_j^b(\t')>&=&\frac{1}{2}\d^{ab}\d_{ij}\D_0(\t,\t') 
 \nn\\
&&~~~~~~~\nn\\
<A^a(\t)A^b(\t')>&=&\frac{1}{2}\d^{ab}\D_0(\t,\t') \nn\\
&&~~~~~~~\nn\\
<X_i^{*\a}(\t)X_j^{\b}(\t')>&=&\frac{1}{2}\d^{\a\b}\d_{ij}
\int_0^{\infty}{\rm d}s\:\D^{\a}(\t,\t',s) \qquad \qquad i,j\neq 1 \nn\\
&&~~~~~~~\nn\\
<A^{*\a}(\t)A^{\b}(\t')>&=&<X_1^{*\a}(\t)X_1^{\b}(\t')>
=\frac{1}{2}\d^{\a\b}\int_0^{\infty}{\rm d}s\:~ {\rm Cosh}(2 \va s) ~\D^{\a}(\t,\t',s) \nn\\
&&~~~~~~~\nn\\
<X_1^{*\a}(\t)A^{\b}(\t')>&=&-<A^{*\a}(\t)X_1^{\b}(\t')>
=- i\frac{1}{2}\d^{\a\b}\int_0^{\infty}{\rm d}s\:~ {\rm Sinh}(2 \va s)~\D^{\a}(\t,\t',s) \nn\\
&&~~~~~~~
\label{propagatorbos}
\eea
while for the ghosts and the fermions we find 
\bea
<\wt{G}^{a}(\t)G^b(\t')>&=&\frac{1}{2}\d^{ab}\D_0(\t,\t') \nn\\
&&~~~~~~~\nn\\
<\tilde{G}^{\a}(\t)G^{\b}(\t')>&=&\frac{1}{2}\d^{\a\b}\int_0^{\infty}{\rm d}s\:\D^{\a}(\t,\t',s)  \nn\\
&&~~~~~~~\nn\\
<\theta^{a T}(\t)\theta^{b}(\t')>&=&\frac{1}{2}\d^{ab}
\pa_{\t}\D_0(\t,\t') \nn\\
&&~~~~~~~\nn\\
<\theta^{*\a T}(\t)\theta^{\b}(\t')>&=&\frac{1}{2}\d^{\a\b}
\bf{\D}^{\a}_F(\t,\t') 
\label{propagatorferm}
\eea
We have defined
\bea
\D_0(\t,\t')&=& \theta(\t-\t')(\t'-\t) \nn\\
&&~~~~~~~\nn\\
\bf{\D}^{\a}_F(\t,\t')& \equiv&  \int_0^{\infty}{\rm d}s\:
\left(-~~\II ~~\frac{\va t}{{\rm Sinh}(\va s)} 
+\g^1\frac{\va T}{{\rm Cosh}(\va s)} \right.  \nn\\
&&~~~~~~~\nn\\
&&~~~~~~~~~~\left. +{\bbar}^{\a}{\rm Cosh}(\va s)+{\bbar}^{\a} \g^1{\rm Sinh}(\va s)\r)
\D^{\a}(\t,\t',s) 
\label{propexplicit}
\eea
where
\bea
\D^{\a}(\t,\t',s) &=&  e^{- (b^{\a})^2 s} 
\sqrt{\frac{\va}{2 \pi {\rm Sinh}(2 s \va)}} 
e^{- \va T^2 {\rm Tanh}(s \va) - \va t^2{\rm Coth}(s \va)}  \nonumber \\
&~&~~\label{defprop}
\eea 
We have introduced also new time variables
\beq
T= \frac{1}{2}(\tau+\tau') ~~~~~~~~~~~~t= \frac{1}{2}(\tau-\tau')
\label{newtime}
\eeq

Once the propagators are known the one-loop contribution to the effective 
action is easily computed \cite{BB}
\beq
\Gamma^{(1)}= -\frac{15}{16} \sum_{\a=1}^{3} \int ~dT~
\frac{(v_\a)^4}{[(b^\a)^2+T^2 (v_\a)^2]^{\frac{7}{2}}}
\label{one-loop}
\eeq

In the next section we present the two-loop calculation.

\section{The two-loop effective action}
The two-loop contributions are evaluated considering two types of diagrams 
involving one four-point vertex, the figure-eight graphs, and 
two three-point vertices, the sunset-type graphs, respectively.
Given our setup, we can distinguish between the ones that lead to genuine
three-body exchanges (one--particle irreducible diagrams for the three--graviton 
scattering in supergravity) and two-body recoil effects (one-particle reducible graphs in supergravity) in a very simple manner. We need collect terms that depend on 
two independent relative velocities on one side, and terms that depend only
on one relative velocity on the other side. 

Thus from the figure-eight diagrams we have a three-body exchange whenever 
the two propagators come in with different masses, while a recoil term is
 produced when the two masses are equal. The presence of a massless 
propagator would lead to a vanishing contribution for this tadpole kind of 
diagrams.

From the sunset-type graphs we have direct three-body contributions
when the three propagators have three distinct masses (three different
relative velocities out of which two are independent). Two-body forces
are present when two masses are equal, being the third one equal to zero
 as required by momentum conservation.
Having decomposed the fields as in (\ref{scompo}) makes it easy to identify
the massless particles which are given simply by the diagonal degrees of 
freedom.

Now we present our results. First we collect all the contributions to the
three-body forces: we follow as much as possible the approach in 
ref. \cite{giappo} so that a direct comparison can be made in a 
straightforward manner. Complete agreement with their findings is shown in
the next subsection. 
Then we concentrate on the two-loop contributions that depend only on one 
relative velocity $v_\a$.  Their complete evaluation is conceptually simple
and algebraically manageable within our approach. We isolate the 
leading order term and prove its consistency with the result 
from two-graviton scattering obtained in \cite{BBP}. This shows without any
further ambiguity that we are dealing with recoil effects.

\subsection{Three-body contributions}
As emphasized above we need compute all the two-loop terms which depend on 
two distinct relative velocities, and this implies that only propagators
associated to off-diagonal degrees of freedom (see (\ref{scompo}))
will enter this part of the calculation. We have to focus on four-point
vertices for the figure-eight diagrams and three point interactions for the
sunset diagrams: they are explicitly given in (\ref{actionbos}), 
(\ref{actiongauge}), (\ref{actionfermion}) and (\ref{actionghost}).
The corresponding propagators can be read from (\ref{propagatorbos}),
(\ref{propagatorferm}) and (\ref{propexplicit}), (\ref{defprop}). In
order to keep the notation compact we write $\Delta^\a(s)\equiv
\Delta^\a(\t,\t',s)$. The full result is given by
\bea
\Gamma &= &-\sum_{\a \neq\b \neq \c} \left[ 
\int dT  ds_2 ds_3 ~ P_0  ~ 
\left( \Delta^\b(s_2) \Delta^\c(s_3) \right)_{|t=0} 
 \right.  \nonumber \\
&&~~~~~ +\int dT dt ds_1 ds_2 ds_3  ~ P_1  ~ (\pa_\tau \pa_{\tau'}  
\Delta^\a(s_1)) \Delta^\b(s_2) \Delta^\c(s_3)   \nonumber \\
&&~~~~~ + \int dT dt ds_1 ds_2 ds_3  ~ P_2  \tau' ~  (\pa_\tau  
\Delta^\a(s_1)) \Delta^\b(s_2) \Delta^\c(s_3)  \nonumber \\
&& ~~~~~ + \int dT dt ds_1 ds_2 ds_3  ~ P_3  ~ \va^2 \tau \tau' ~ 
\Delta^\a(s_1) \Delta^\b(s_2) \Delta^\c(s_3)  \nonumber \\ 
&& ~~~~~ + \int dT dt ds_1 ds_2 ds_3  ~ P_4 ~  \va v_\b \tau 
\tau' ~ \Delta^\a(s_1) \Delta^\b(s_2) \Delta^\c(s_3)  \nonumber \\ 
&&~~~~~ + \int dT dt ds_1 ds_2 ds_3  ~ P_5  ~ (b^\a)^2  ~ \Delta^\a(s_1) 
\Delta^\b(s_2) \Delta^\c(s_3)  \nonumber \\
 &&~~~~~ + \left. \int dT dt ds_1 ds_2 ds_3  ~ P_6  ~ b^\a b^\b  ~ 
\Delta^\a(s_1) \Delta^\b(s_2) \Delta^\c(s_3)\right]
\label{3body}
\eea
where the complete expressions of the various $P$'s are given in Appendix B.

At this point we perform manipulations similar to the ones introduced
in ref. \cite{giappo}: this will lead to an easy comparison of our results
and theirs. Since 
\beq
b^\a + b^\b + b^\c = 0 \qquad\qquad v_\a + v_\b + v_\c = 0 
\qquad \qquad \a \neq \b \neq \c
\eeq
we have
\beq
\sum_{\a \neq\b \neq \c}b^\a  b^\b f({\a,\b,\c}) =\sum_{\a \neq\b \neq \c}
\frac{1}{2} (b^\a)^2 \left[ f({\c,\b,\a}) - f({\a,\b,\c}) - f({\a,\c,\b}) 
\right]
 \label{iden0}
\eeq
The identity in (\ref{iden0}) allows to combine
 $P_6$ and $P_5$: we rename the resulting expression 
$P^{new}_5$. In the same way, using again (\ref{iden0}) we combine 
$P_4$ and $P_3$ and rename their sum $P^{new}_3$.
Moreover as in \cite{giappo} we can write
\bea
&& \int dT  ds_2 ds_3 P \left( \Delta^\b(s_2) \Delta^\c(s_3) 
\right)_{|t=0} =
\int dT dt ds_2 ds_3 P \delta(t) \Delta^\b(s_2) \Delta^\c(s_3) 
\nonumber \\
&&~~~~~~~~= - \int dT dt ds_1 ds_2 ds_3 P \left( \pa_{s_1} \Delta^\a(s_1) 
\right) \Delta^\b(s_2) \Delta^\c(s_3)
\label{iden1}
\eea
and use the following relation
\bea
(b^\a)^2 \D^\a(s_1) &=&  -\pa_{s_1} \Delta^\a(s_1)+ \left(- \va^2 T^2
 - \va^2 t^2 + \va^2 T^2 \frac{\sha^2}{\cha^2}  \right. \nonumber \\ 
&&~~~ \left.+\va^2 t^2 \frac{\cha^2}
{\sha^2} - \frac{1}{2} \va \frac{\sha}{\cha}  - \frac{1}{2} \va 
\frac{\cha}{\sha}\right)\D^\a(s_1) \nn\\
&&~~~~~~~
\label{iden2}
\eea 
It is rather straightforward to show that, with the help of 
(\ref{iden0}), (\ref{iden1}) and (\ref{iden2}), the answer in (\ref{3body})
can be rearranged as a sum of two contributions
\beq
\Gamma=\Gamma_V+\Gamma_Y
\label{V+Y}
\eeq
Here
\bea
\Gamma_V &=& -\sum_{\b \neq \c} \int dT  ds_2 ds_3 
\left({P_5^{new}}_{|{s_1}=0} + P_0 \right)  \left( \Delta^\b(s_2) 
\Delta^\c(s_3) \right)_{|t=0}\nn\\
 &=&-\sum_{\b \neq \c} \int dT  ds_2 ds_3 ~ 128~ 
{\rm Sinh}^3(\frac{s_2 v_\b}{2}) ~  {\rm Sinh}^3(\frac{s_3 v_\c}{2}) 
\times \nonumber \\
&~&~\left ( 2  {\rm Cosh}(\frac{s_2 v_\b}{2})  
{\rm Cosh}(\frac{s_3 v_\c}{2})-  {\rm Sinh}(\frac{s_2 v_\b}{2})   
{\rm Sinh}(\frac{s_3 v_\c}{2}) \right)  \Delta^\b(s_2) \Delta^\c(s_3)  
\nonumber \\
&~&~ \label{gammaV}
\eea
This result exactly reproduces the corresponding term obtained in 
ref. \cite{giappo}.

The second term in (\ref{V+Y}) is given by
\bea
\Gamma_Y &=& - \sum_{\a \neq \b \neq \c} \int dT dt ds_1 ds_2 ds_3
~\left( \frac{\va}{ {\rm Sinh}(2 s_1 \va)} + \va^2 T^2 {\rm Tanh}^2(s_1 \va) -
 \va^2 t^2 {\rm Coth}^2(s_1 \va) \right) \times \nonumber \\
&&~~~~~~~~~~~~~~~~~~~~~~~~~~~~~~~~~~~~~~~~~~~~~~~~~~P_1   \Delta^\a(s_1) \Delta^\b(s_2) \Delta^\c(s_3)   \nonumber \\
&& ~~~ \nonumber \\  
&& ~~~~+ \int dT dt ds_1 ds_2 ds_3  ~\left (- \va T^2 {\rm Tanh}(s_1 \va) +
 \va t^2 {\rm Coth}(s_1 \va) \right) \times \nonumber \\  
&&~~~~~~~~~~~~~~~~~~~~~~~~~~~~~~~~~~~~~~~~~~~~~~~~~~
 P_2  \Delta^\a(s_1) 
\Delta^\b(s_2) \Delta^\c(s_3)  \nonumber \\
&& ~~~ \nonumber \\  
&& ~~~~+\int dT dt ds_1 ds_2 ds_3  ~ P_3^{new}  ~ \va^2 \tau \tau' ~ \Delta^\a(s_1) 
\Delta^\b(s_2) \Delta^\c(s_3)  \nonumber \\ 
&& ~~~~+\int dT dt ds_1 ds_2 ds_3  ~  \left( \pa_{s_1} P_5^{new}  ~ - \left(\va^2 T^2 
+ \va^2 t^2 \right) ~ P_5^{new}\right)  ~ \D^\a(s_1) \Delta^\b(s_2) \Delta^\c(s_3)  
\nonumber \\
&& ~~~ \nonumber \\
&& ~~~~+\int dT dt ds_1 ds_2 ds_3 \left( \va^2 T^2 \frac{\sha^2}{\cha^2} + 
\va^2 t^2 \frac{\cha^2}{\sha^2}\right) \times \nn\\
&&~~~~~~~~~~~~~~~~~~~~~~~~~~~~~~~~~~~~~~~~~~~~~~~~~~P_5^{new} \D^\a(s_1) \Delta^\b(s_2) 
\Delta^\c(s_3)    \nonumber \\
&& ~~~ \nonumber \\
&& ~~~~+\int dT dt ds_1 ds_2 ds_3  \left( - \frac{1}{2} \va 
\frac{\sha}{\cha}  - \frac{1}{2} \va \frac{\cha}{\sha}\right) \times \nn\\
&&~~~~~~~~~~~~~~~~~~~~~~~~~~~~~~~~~~~~~~~~~~~~~~~~~~
P_5^{new} \D^\a(s_1) \Delta^\b(s_2) \Delta^\c(s_3) \nonumber \\
&& ~~~~ 
\label{gammaY}
\eea
Further
simplifications are introduced using the identity
\beq
\int_{-\infty}^{+\infty} dr~r^2 e^{-Qr^2}= 
\int_{-\infty}^{+\infty} dr~\frac{1}{2Q} ~ e^{-Qr^2}
\label{idint}
\eeq
Indeed, recalling the expression of $\D^\a$ in (\ref{defprop}), 
the above identity allows to perform the following substitutions in the
integrals in $T$ and $t$:
\bea
T^2 \D^\a(s_1)\D^\b(s_2)\D^\gamma(s_3)  &\rightarrow& 
\frac{\D^\a(s_1)\D^\b(s_2)\D^\gamma(s_3)}{2 \left( \va {\rm Tanh}(\va s_1) + v_\beta 
{\rm Tanh}(v_\beta s_2)+  v_\gamma {\rm Tanh}(v_\gamma s_3)\right)} 
\nonumber \\
&&\nonumber \\
t^2 \D^\a(s_1)\D^\b(s_2)\D^\gamma(s_3)
&\rightarrow& \frac{\D^\a(s_1)\D^\b(s_2)\D^\gamma(s_3)}{2 \left( \va {\rm Coth}(\va s_1) + v_\beta 
{\rm Coth}(v_\beta s_2) +v_\gamma {\rm Coth}
(v_\gamma s_3)\right)}\nonumber \\ 
&&~~\label{substit} 
\eea
Finally in (\ref{gammaY})
we can expand the various coefficients in front of the common factor $\D^\a(s_1) \Delta^\b(s_2) \Delta^\c(s_3)$ in powers of $v$ and find the first non vanishing contribution at order $v^8$
\bea
\Gamma_Y &=& - \sum_{\a \neq \b \neq \c} \int dT dt ds_1 ds_2 ds_3 ~
\frac{1}{18}
(s_1 \va - s_2 v_\b)^2 (s_1 \va - s_3 v_\c)^2 (s_2 v_\b - s_3 v_\c)^2
  \nonumber \\
&&~~~~~~~~~~~~~~~~ (s_1 \va^2 + s_2 v_\b^2 + s_3 v_\c^2) \D^\a(s_1) 
\Delta^\b(s_2) \Delta^\c(s_3)
\label{v8term}
\eea
We note that in ref. \cite{giappo} the corresponding leading contribution is
at order $v^6$. With our choice of the background it is easy to check that
the order $v^6$ term vanishes identically since all the relative 
velocities are parallel to each other. The order $v^8$ contribution is 
contained in the complete answer given in \cite{giappo}, but algebraically
too difficult to be reconstructed. In our case it can be evaluated rather easily
since it is the first non vanishing contribution.

Now we turn to the calculation of the recoil effects.

\subsection{Two-body recoil contributions}

We consider here all the diagrams not computed in the previous section,
i.e. figure-eight graphs with the two propagators carrying the
same mass, and sunset-type graphs with two propagators of equal masses 
and the third one massless.  Looking at the spectrum of the various 
particles it is clear that these contributions depend on a single relative
 velocity $v_\a$ and are thus candidates to represent supegravity
two-body recoil effects. In this section we determine them  exactly
and show that the above interpretation is indeed confirmed.

The complete  amplitude can be written 
(cfr. (\ref{3body})): 
\bea
\Gamma_{recoil}& =&-\sum_{\a }\left[\int dT  ds_1 ds_2 ~ Q_0~  \left( 
\Delta^\a(s_2) \Delta^\a(s_3) \right)_{|t=0}  \right.
\nonumber \\
&&~~~~~ +\int dT dt ds_1 ds_2  ~ Q_1~  \Delta_0 (\pa_\tau \pa_{\tau'}  
\Delta^\a(s_2)) \Delta^\a(s_1)    \nonumber \\
&&~~~~~ +\int dT dt ds_1 ds_2 ~  Q_2~  \Delta_0 (\pa_\tau 
\Delta^\a(s_1)) \  (\pa_{\tau'} \Delta^\a(s_2))  \nonumber \\
&& ~~~~~+\int dT dt ds_1 ds_2 ~  Q_3~  \Delta_0 (\va \tau' \pa_\tau  
\Delta^\a(s_2)) \Delta^\a(s_2)   \nonumber \\
&&~~~~~+ \int dT dt ds_1 ds_2~  Q_4~  \va^2 \tau \tau' ~ \Delta_0 
\Delta^\a(s_1) \Delta^\a(s_2)   \nonumber \\  
&&~~~~~+\int dT dt ds_1 ds_2 ~  Q_5~  (b^\a)^2~  \Delta_0 
\Delta^\a(s_1) \Delta^\a(s_2)   \nonumber \\
&& ~~~~~+\int dT dt ds_1 ds_2 ~  Q_6~  \va^2 T^2~  \Delta_0 \Delta^\a(s_1) 
\Delta^\a(s_2)   \nonumber \\
&&~~~~~+\int dT dt ds_1 ds_2 ~  Q_7~  \va^2 t^2~  \Delta_0 \Delta^\a(s_1) 
\Delta^\a(s_2)   \nonumber \\
&&~~~~~+ \left. \int  dT dt ds_1 ds_2 ~  Q_8 ~  \Delta_0 \Delta^\a(s_1) 
\Delta^\a(s_2) \right]
\label{2body}
\eea
where the expressions of the $Q$ coefficients are given in Appendix C. If we 
compare (\ref{2body}) with the corresponding three-body exchange 
contributions in (\ref{3body}) we notice that in the present case each term 
contains only one $v_\a$. As before the figure-eight terms are the ones 
in which $Q_0$ appears; all the rest comes from sunset graphs with one
massless propagator, $\D_0$.

As we have done for the three-body calculation we can  simplify the 
form of the result. We use the following relation
\bea
&& \int dT  ds_1 ds_2 ~Q_0~  \left( \Delta^\a(s_1) \Delta^\a(s_2) 
\right)_{|t=0} =
\int dT dt ds_1 ds_2 ~Q_0~ \delta(t) \Delta^\a(s_1) \Delta^\a(s_2) 
\nonumber \\
&&~~= \int dT dt  ds_1 ds_1 ~Q_0 ~\left( \pa_{\t'} \pa_\t \Delta_0 \right) 
\Delta^\a(s_1) \Delta^\a(s_2) \nonumber \\
&& ~~=\int dT dt  ds_1 ds_1 ~Q_0~ \Delta_0 \Bigl(  \pa_{\t'} \pa_\t 
\Delta^\a(s_1) \Delta^\a(s_2) + 2 \pa_{\t'} \Delta^\a(s_1) \pa_\t 
\Delta^\a(s_2)  \nn\\
&&~~~~~~~~~~~~~ +\Delta^\a(s_1)  \pa_{\t'} \pa_\t \Delta^\a(s_2)
 \Bigl) 
\label{iden1n}
\eea
Given the expressions of the propagators in (\ref{defprop}), we are able to perform the derivatives 
in (\ref{2body}) explicitly.
In addition, in complete analogy with what we have done in (\ref{idint}) and
in (\ref{substit}), we can
show that, since we are integrating over $T$ and $t$ the following 
substitutions are allowed
\bea
T^2 \D_0 \D^\a(s_1)\D^\a(s_2)
&\longrightarrow& \frac{1}{2 \va \left( {\rm Tanh}(\va s_1) +  
{\rm Tanh}(\va s_2)\right)} \D_0 \D^\a(s_1)\D^\a(s_2)\nonumber \\
t^2 \D_0 \D^\a(s_1)\D^\a(s_2)&\longrightarrow& 
\frac{1}{ \va \left( {\rm Coth}(\va s_1) +  
{\rm Coth}(\va s_2)\right)}\D_0 \D^\a(s_1)\D^\a(s_2) \nonumber \\
&&~~~\label{substit2} 
\eea
The above substitutions lead to
\beq
\Gamma_{recoil} =- \sum_{\a }\left[ \int dT dt ds_1 ds_2 
\Bigl((b^\a)^2S_1+S_2\Bigl)\D_0 \D^\a(s_1)\D^\a (s_2)\right]
\label{recpart}
\eeq
where $S_1$ and $S_2$ are given by
\bea
S_1&=& 40 + 2 ~{\rm Cosh}(2 s_1 \va) + 32 ~ {\rm Cosh}((s_1 - s_2) \va) + 2~  {\rm Cosh}(2 s_2 \va)   \nonumber\\
&&~~ +96~  {\rm Cosh}((s_1 + s_2) \va) + 8 ~ {\rm Cosh}(2 (s_1 + s_2) \va)
\nonumber\\
S_2&=& \frac{1}{4} \Bigl[-99~ \va + 128 ~\va  {\rm Cosh}(s_1 \va) + 37 ~\va  {\rm Cosh}(2 s_1 \va) - 3~ \va  {\rm Cosh}(4 s_1 \va)   \nonumber\\
&&~~  -384~ \va  {\rm Cosh}((s_1 - s_2) \va) - 48 ~ \va  {\rm Cosh}(2 (s_1 - s_2) \va) + 128~  \va  {\rm Cosh}(s_2 \va)   \nonumber\\
&&~~ +37 ~ \va  {\rm Cosh}(2 s_2 \va) - 3~  \va  {\rm Cosh}(4 s_2 \va) + 384 ~ \va  {\rm Cosh}((s_1 + s_2) \va)  \nonumber\\
&&~~ +92~  \va  {\rm Cosh}(2 (s_1 + s_2) \va) +  \va  {\rm Cosh}(4 (s_1 + s_2) \va) - 
 128 ~ \va  {\rm Cosh}((2 s_1 + s_2) \va)  \nonumber\\
&&~~ -7 ~ \va  {\rm Cosh}(2 (2 s_1 + s_2) \va) -  128~  \va  {\rm Cosh}((s_1 + 2 s_2) \va)  \nonumber\\
&&~~ - 7 ~ \va  {\rm Cosh}(2 (s_1 + 2 s_2) \va) \Bigl] {\rm Cosech}(s_1 \va) 
  {\rm Cosech}(s_2 \va)  {\rm Cosech}((s_1 + s_2) \va) \nonumber\\
&&~~ 
\label{Spoly}
\eea
At this stage we integrate on $t$ and $T$ explicitly and 
 expand the result in power of $\va$ up to order $\va^5$. We find
\beq
\Gamma_{recoil} =-\sum_{\a }\left[ \int  ds_1 ds_2 
\frac{\Bigl((b^\a)^2 V_1+V_2\Bigl)~ e^{-(b^\a)^2(s_1+s_2)}}{2\p \va \sqrt{s_1s_2} \sqrt{s_1+s_2}}
\right]
\label{recexp}
\eeq
where
\bea
V_1&=&  \left( 75600 + 22680~ s_1^2  \va^2  + 2520~ s_1 s_2 \va^2  + 22680 ~s_2^2  \va^2  + 630~ s_1^4  \va^4  -  2100~ s_1^3  s_2 \va^4  \right.  \nonumber\\
&&~~ +10290~ s_1^2  s_2^2  \va^4  - 2100~ s_1 s_2^3  \va^4  + 630~ s_2^4  \va^4  +  139~ s_1^6  \va^6  + 761~ s_1^5  s_2 \va^6   \nonumber\\
&&~~ +\left. 1794~ s_1^4  s_2^2  \va^6  - 947 ~s_1^3  s_2^3  \va^6  + 1794~ s_1^2  s_2^4  \va^6  + 761~ s_1 s_2^5  \va^6  + 139~ s_2^6  \va^6 \right) \frac{s_1 s_2}{840 ~(s_1 + s_2)}
 \nonumber\\
V_2&=& \left(-75600~ s_1^2  + 302400~ s_1 s_2 - 75600~ s_2^2  - 17640~ s_1^4   \va^2  - 27720~ s_1^3  s_2 \va^2    \right.  \nonumber\\
&&~~-277200~ s_1^2  s_2^2  \va^2  - 27720~ s_1 s_2^3  \va^2  -17640~ s_2^4  \va^2  - 6510~ s_1^6  \va^4  + 2940~ s_1^5  s_2 \va^4   \nonumber\\
&&~~ +68040~ s_1^4  s_2^2  \va^4  + 211680~ s_1^3  s_2^3  \va^4  + 68040~ s_1^2  s_2^4  \va^4  + 2940~ s_1 s_2^5  \va^4  -6510~ s_2^6  \va^4   \nonumber\\ 
&&~~ -97~ s_1^8  \va^6  + 2065~ s_1^7  s_2 \va^6  - 7695~ s_1^6  s_2^2  \va^6  - 51738~ s_1^5  s_2^3  \va^6  - 103508~ s_1^4  s_2^4  \va^6     \nonumber\\ 
&&~~ -\left. 51738~ s_1^3  s_2^5  \va^6  -7695~ s_1^2  s_2^6  \va^6  + 2065~ s_1 s_2^7  \va^6  - 97~ s_2^8  \va^6 \right)\frac{1}{3360~ (s_1 + s_2)^2 }
\label{Vpoly}
\eea
Finally the integrations on $s_1$ and $s_2$ can be performed using the 
following relation
\beq
\int ds_1~ds_2 \frac{s_1^\ell ~s_2^m ~e^{-(b^\a)^2(s_1+s_2)}}{(s_1+s_2)^n
\sqrt{s_1s_2}\sqrt{s_1+s_2}}
=\frac{1}{(b^\a)^{2 \ell+2 m -2 n+1}} \b\left(\ell+\frac{1}{2}, m + \frac{1}{2}\right)\G\left(\ell+m -n+\frac{1}{2}\right)
\label{formula}
\eeq
Indeed with the help of (\ref{formula}) it is easy to show that
\bea
&& \int  ds_1 ds_2 ~ ~ 
\frac{~ (b^\a)^2 ~ V_1 ~ e^{-(b^\a)^2(s_1+s_2)}}{2\p \va ~ \sqrt{s_1s_2} ~ \sqrt{s_1+s_2}} \nonumber \\
&&~= \frac{45 \pi~ (4194304 ~ (b^\a)^{12}   + 3047424 ~ (b^\a)^8  \va^2  + 1241856 ~ (b^\a)^4  \va^4  + 11390093 ~ \va^6 )}{67108864 ~ (b^\a)^{13}  ~  \va}\nonumber \\
&&~
\label{1int}
\eea
and
\bea
&&\int  ds_1 ds_2 ~ ~ 
\frac{ V_2 ~ e^{-(b^\a)^2(s_1+s_2)}}{2\p \va ~ \sqrt{s_1s_2} ~ \sqrt{s_1+s_2}} \nonumber \\
&&~= 
-\frac{45 \pi (4194304 ~ (b^\a)^{12}   + 3047424 ~ (b^\a)^8  \va^2  + 1241856 ~ (b^\a)^4  \va^4  + 10207373 ~ \va^6 )}{67108864 ~ (b^\a)^{13} ~   \va} \nonumber \\
&&~
\label{2int}
\eea
Summing the two contributions in (\ref{1int}) and (\ref{2int}) we obtain
\beq
\Gamma_{recoil}= - \sum_\a \frac{51975}{65536} \pi \frac{\va^5}{(b^\a)^{13}}
 \label{final1}
\eeq
We write (\ref{final1}) in the form
\beq
\Gamma_{recoil}= -\sum_\a \int dT ~\frac{225}{64} ~\frac{\va^6}{((b^\a)^2 + \va^2 T^2)^{7}} 
\label{final2}
\eeq 
The expression in (\ref{final2}) can be directly compared with 
the result in ref. \cite{BBP} where the two graviton scattering was
analyzed: we find that numbers perfectly match. The expression in (\ref{recpart}) with $\Gamma_V$ in (\ref{gammaV}) and $\Gamma_Y$ in (\ref{gammaY}) determines the {\em full} two--loop contribution to the three D--particle effective action.
 
\section{Conclusions}
In this paper we have presented a complete two-loop calculation in $M$-atrix
theory. We have considered the gauge group $U(3)$, thus focusing on the 
description of three $D~0$-branes. In order to maintain the algebraic 
complexity of the computation under control, we have made a specific choice 
of the classical background, i.e. we have restricted our attention to the 
case of three $D$-particles whose relative velocities are
parallel and orthogonal to the corresponding relative displacements. 
First we have confirmed the results obtained in \cite{giappo}, showing 
that the predictions from {\em one-particle irreducible} tree diagrams in supergravity are in agreement with  two-loop 
 contributions from $M$-atrix theory.
Then we have considered those two-loop  contributions which depend only on one relative velocity.
The novelty of our result resides primarily in this part of our work. 
We have shown that this new type of terms gives a total sum which is  exactly consistent with what is obtained  from a
two-loop calculation of two $D$-particle scattering \cite{BBP}. Therefore
we can safely conclude that we have determined those effects
which amount to {\em one-particle reducible} tree level diagrams in supergravity. 
Once more one finds impressive evidence that eleven-dimensional
supergravity compactified in a null direction is in direct correspondence
with $M$-atrix theory results for finite $N$.

\medskip
\section*{Acknowledgments.}

\noindent 

This work was
supported by the European Commission TMR program
ERBFMRX-CT96-0045, in which A. R., N. T. and D. Z. are associated 
to the University of Torino. 
\newpage

\appendix
\section{Lie algebra for $SU(3)$}
We have chosen the following Cartan basis for the Lie algebra of $SU(3)$
\bea
H^1=\frac{1}{\sqrt{6}}\left( \matrix{ 1 & 0 & 0 \cr 0 & 1 & 0 
\cr 0 & 0 & -2 }\right)
&& H^2=\frac{1}{\sqrt{2}} \left(\matrix{1 & 0 & 0 \cr 0 & -1 & 0 \cr 
0 & 0 & 0 } \right)\nonumber\\
~~~~~~~&&~~~~~~~~\nonumber\\
~~~~~~~&&~~~~~~~~\nonumber\\
E_{\a^1} = \left(\matrix{ 0 & 0 & 0 \cr 0 & 0 & 1 \cr 0 & 0 & 0 }\right)
  && E_{-\a^1} =\left( \matrix{0 & 0 & 0 \cr 0 & 0 & 0 \cr 0 & 1 & 0 }
\right)\nonumber\\
~~~~~~~&&~~~~~~~~\nonumber\\
~~~~~~~&&~~~~~~~~\nonumber\\
E_{\a^2} = \left(\matrix{ 0 & 0 & 0 \cr 0 & 0 & 0 \cr 1 & 0 & 0 }\right)
 && E_{-\a^2} = \left(\matrix{0 & 0 & 1 \cr 0 & 0 & 0 \cr 0 & 0 & 0 }\right)
\nonumber\\
~~~~~~~&&~~~~~~~~\nonumber\\
~~~~~~~&&~~~~~~~~\nonumber\\
E_{\a^3} = \left(\matrix{ 0 & 1 & 0 \cr 0 & 0 & 0 \cr 0 & 0 & 0 }\right)
 && E_{-\a^3} = \left(\matrix{ 0 & 0 & 0 \cr 1 & 0 & 0 \cr 0 & 0 & 0}\right)
\label{generators}
\eea
with root vectors
\beq 
\a^1=\left(\sqrt{\frac{3}{2}},-\frac{1}{\sqrt{2}}\right),\qquad\qquad 
  \a^2=\left(-\sqrt{\frac{3}{2}},-\frac{1}{\sqrt{2}}\right),\qquad \qquad
  \a^3=\left(0,\sqrt{2}\right)  
\label{roots}
\eeq
normalized as
\beq
\a\cdot \b= \left\{ \matrix{ 2 ~~& if~~ & \a=\b \cr -1 ~~& if ~~
& \a \neq \b }\right.
\qquad \qquad\qquad \a^1+\a^2+\a^3=0    
\eeq

The generators satisfy the following commutation relations
\bea
\left[ H^i , E_{\a} \right] = \a_i E_{\a} & \qquad & \left[ H^i ,  
E_{-\a} \right] = -\a_i E_{-\a} \nonumber \\
\left[ E_{\a^1} , E_{\a^2} \right] = E_{-\a^3} & \qquad & \left[ E_{-\a^1} ,
 E_{-\a^2} \right] = -E_{\a^3} \nonumber \\
\left[ E_{\a^2} , E_{\a^3} \right] = E_{-\a^1} & \qquad & \left[ E_{-\a^2} ,
 E_{-\a^3} \right] = -E_{\a^1} \nonumber \\
\left[ E_{\a^3} , E_{\a^1} \right] = E_{-\a^2} & \qquad & \left[ E_{-\a^3} ,
 E_{-\a^1} \right] = -E_{\a^2} \nonumber 
\eea
\bea
& \left[ E_{\a^1} , E_{-\a^1} \right] = \a^1 \cdot H & \nonumber \\
& \left[ E_{\a^2} , E_{-\a^2} \right] = \a^2 \cdot H & \nonumber \\
& \left[ E_{\a^3} , E_{-\a^3} \right] = \a^3 \cdot H & \nonumber \\
& \left[ H^i , H^j \right] = 0 &
\label{algebra}
\eea

\section{Coefficients for three-body forces}
We list here the expressions of $P_0,\ldots,P_6$ which appear into the two--loop effective action for direct three--body exchanges.
\bea
 P_0 &=& - (45 + 18 \shc^2 +18 \shb^2 + 4 \shc^2 \shb^2  \nonumber \\ 
&&~~~~ - 12 \shb \shc \chb \chc) 
\eea
\bea
P_1&=& 4 ( 1+ 2 \shb^2) + 4 ( 1+ 2 \shc^2)  \nonumber \\
&& +( 1+ 2 \sha^2)( 1+ 2 \shb^2)( 1+ 2 \shc^2)  \nonumber \\
&&- 4 \shb \shc \chb \chc ( 1+ 2 \sha^2)  \nonumber \\
&&- 6 \shb \chb \shc \chc (1+ 2 \sha^2)  \nonumber \\
~~&& +2 \sha \cha \shc \chc (1+ 2 \shb^2)  \nonumber \\
&&+2 \sha \cha \shb \chb (1+ 2 \shc^2) \nonumber \\
&& +1/2 (1 + 2 \sha^2) (1+ 2 \shb^2) (1+ 2 \shc^2)  \nonumber \\
&&- 7/2 (1 + 2 \sha^2) + 7/2 (1+ 2 \shb^2) + 7/2 (1+ 2 \shc^2) \nonumber \\
~~&&+32 ( \cha \chb + \sha \shb)   \nonumber \\
&&+32 (\cha \chc + \sha \shc)  \nonumber \\
&&-32 ( \chb \chc + \shb \shc)
\eea
\bea
P_2&=&  \va (- 16 \sha \shb \shc \cha \chb \chc  \nonumber \\
&&~+4 \sha \cha (1+ 2 \shb^2)(1+ 2 \shc^2) \nonumber \\
~~&&~+16 \shc \chc + 16 \shb \chb ) \nonumber \\
&& +\vb (- 4 \sha \cha (1+ 2 \shb^2)(1+ 2 \shc^2)  \nonumber \\
&&~- 4 \shb \chb (1+ 2 \sha^2)(1+ 2 \shc^2)\nonumber \\
~~&&~+ 4 \shc \chc (1+ 2 \shb^2)(1+ 2 \sha^2) \nonumber \\
&& ~+16 \sha \shb \shc \cha \chb \chc  \nonumber \\
&&~- 32 \shc \chc + 4 \shc \chc  \nonumber \\
&&~- 128 (\cha \shb + \chb \sha) )
\eea
\bea
P_3&=& ( 1+ 2 \sha^2)( 1+ 2 \shb^2)( 1+ 2 \shc^2) \nonumber \\
&&- 4 ( 1+ 2 \sha^2) \shb \shc \chb \chc  \nonumber \\
&&+ 4 (1 + 2 \shb^2) + 4 ( 1+ 2 \shc^2)
\eea
\bea
P_4&=& 4 ( 1+ 2 \sha^2) \shb \shc \chb \chc  \nonumber \\
&&+4 ( 1+ 2 \shb^2) \sha \shc \cha \chc  \nonumber \\
&&-4 ( 1+ 2 \shc^2) \shb \sha \chb \cha \nonumber \\
~~&&-  ( 1+ 2 \sha^2)( 1+ 2 \shb^2)( 1+ 2 \shc^2)  
 \nonumber \\
&& - 8 (1 + 2 \shc^2)+(1 + 2 \shc^2) \nonumber \\
&& - 64 ( \sha \shb + \cha \chb) 
\eea
\bea
P_5&=&  7 +  (1+ 2 \sha^2 ) (1 + 2 \shb^2)~~~~~\qquad \qquad
 \qquad \qquad \nonumber \\
&& +(1+ 2 \sha^2 ) 
(1 + 2 \shc^2) \nonumber \\
&& - 4 \shb \sha \chb \cha  \nonumber \\
&&- 4 \sha \shc \cha \chc
\eea
\bea
P_6&=& -( 1+ 2 \sha^2)( 1+ 2 \shb^2) -6  \nonumber \\
&&- (1+ 2 \sha^2 ) 
(1 + 2 \shb^2)  \nonumber \\
&& +8 \sha \shb \cha \chb +48 \sha \shb  \nonumber \\
&&- 48 \cha \chb \nonumber \\
~~&&-16 (1+2 \shc^2) (\sha \shb + \cha \chb)  \nonumber \\
&&+32 \chc \shc (\sha \chb +\cha \shb) \nonumber \\
&&
\eea 

\section{Coefficients for two-body forces}
We list here the expressions of $Q_0,\ldots,Q_8$ which appear into the two--loop effective action for two--body recoil exchanges.
\bea
Q_0 &=& -90 - 36 \shd^2 - 36 \shu^2 -8 \shu^2 \shd^2  \nonumber \\
&& ~~~~~~ - 24 \shu \shd \chu \chd
\eea
\bea
Q_1 &=&   16 + 12 (1+ 2 \shd^2) (1+ 2 \shu^2)  \nonumber \\ 
&& ~~~~~~+32  \shu \shd \chu \chd  \nonumber \\
&& ~~~~~~+60  (1+ 2 \shu^2) + 16  (1+ 2 \shd^2) 
\eea
\bea
Q_2 &=& - 14 + 6 (1+ 2 \shd^2) (1+ 2 \shu^2)  \nonumber \\
&& ~~~~~~+40  \shu \shd \chu \chd  \nonumber \\
&& ~~~~~~+30 (1+ 2 \shu^2) + 30 (1+ 2 \shd^2)  
\eea
\bea
Q_3 &=& 2 \left[ 4  \shu \chu (1+ 2 \shd^2)\right. \nonumber \\
&& ~~~~~~+20   \shd \chd (1+ 2 \shu^2) \nonumber \\
&& ~~~~~~-\left. 60  \shu \chu - 28  \shd \chd \right]
\eea
\bea
Q_4  &=&   30 + 6 (1+ 2 \shd^2) (1+ 2 \shu^2)  \nonumber \\
&& ~~~~~~-8 \shu \shd \chu \chd  \nonumber \\
&& ~~~~~~+8 (1+ 2 \shu^2) + 8 (1+ 2 \shd^2) 
\eea
\bea
 Q_5  &=&  40 + 8 (1+ 2 \shd^2) (1+ 2 \shu^2) \nonumber \\
&& ~~~~~~+32 \shu \shd \chu \chd \nonumber \\
&& ~~~~~~+2 (1+ 2 \shu^2) + 2 (1+ 2 \shd) \nonumber \\
&& ~~~~~~+128  \chu \chd + 64 \shu \shd
\eea
\bea
 Q_6  &=& \frac{128}{ \chu \chd} + \frac{128 \shd^2}{\chu} \nonumber \\
&& ~~~~~~~~\nonumber \\
&& ~~~~~~~~~~~~~~~~~~~~+128 \frac{\shu \shd \chu}{\chd^2} 
\eea
 \bea
 Q_7  &=&  \frac{128}{ \shu \shd} -\frac{256 \chd}{ \shu \shd}  -\frac{256 \chd}{ \shd^2 } \nonumber \\
&& ~~~~~~~~~ 
\eea
\bea
Q_8 &=&  \frac{64 \va}{\shd} + \frac{64 \va}{\shu} -
\frac{32  \va}{\chd} \shu \chu \nonumber \\
&& ~~~~~~~~\nonumber \\
&& ~~~~~~~~-\frac{32  \va}{\chu} \shd \chd 
\eea

\end{document}